\documentclass[reprint,amsmath,amssymb,aps,prb,noeprint,twoside]{revtex4-2}

\usepackage{graphicx}
\usepackage{dcolumn}
\usepackage[caption=false]{subfig}

\usepackage{bm}
\usepackage{bbm}
\usepackage{mathrsfs}
\newcommand{\tnn}{t}
\newcommand{\tnnn}{t'}

\usepackage{setspace}
\usepackage{hyperref}
\hypersetup{colorlinks=true,
linkcolor=blue,
filecolor=blue,
urlcolor=blue,
citecolor=blue,
pdftitle={Interpretable AI Analysis of Hubbard Model},
pdfauthor={Changkai Zhang}}

\usepackage{geometry}
\newgeometry{left=0.6in,right=0.6in,top=0.9in,bottom=0.7in}
\usepackage{fancyhdr}

\fancypagestyle{first}{%
    \lhead{}\rhead{}
    \chead{\large PREPRINT FOR PHYSICAL REVIEW (2025)}
}

\usepackage{datetime}

\usepackage{times}
\usepackage[utf8]{inputenc}  
\usepackage{CJKutf8}
\newcommand{\zh}[1]{%
\begin{CJK}{UTF8}{gbsn}#1\end{CJK}%
}

\newlength\mylen
\setcitestyle{citesep={,\kern-\mylen\kern1pt}}
\DeclareUnicodeCharacter{2212}{-}

\begin{document}

\preprint{APS/123-QED}


\title{\vspace*{12pt}Finite-Temperature Study of the Hubbard Model via Enhanced\\[1pt] Exponential Tensor Renormalization Group\vspace{3pt}}%

\author{Changkai Zhang (\zh{张昌凯})}
\author{Jan von Delft\vspace{2pt}}
\affiliation{Arnold Sommerfeld Center for Theoretical Physics,
Center for NanoScience,~and Munich Center for Quantum Science and Technology,\\
Ludwig-Maximilians-Universität München, 80333 Munich, Germany}
\vspace{2pt}


\begin{abstract}
\setstretch{1.08}
The two-dimensional (2D) Hubbard model has long attracted interest for its rich phase diagram and its relevance to high-$T_c$ superconductivity. However, reliable finite-temperature studies remain challenging due to the exponential complexity of many-body interactions. Here, we introduce an enhanced 1s\textsuperscript{+} eXponential Tensor Renormalization Group algorithm that enables efficient finite-temperature simulations of the 2D Hubbard model. By exploring an expanded space, our approach achieves two-site update accuracy at the computational cost of a one-site update, and delivers up to 50\% acceleration for Hubbard-like systems, which enables simulations down to $T\!\approx\!0.004t$. This advance permits a direct investigation of superconducting order over a wide temperature range and facilitates a comparison with zero-temperature infinite Projected Entangled Pair State simulations. Finally, we compile a comprehensive dataset of snapshots spanning the relevant region of the phase diagram, providing a valuable reference for Artificial Intelligence-driven analyses of the Hubbard model and a comparison with cold-atom experiments.
\vspace{24pt}
\end{abstract}

\maketitle

\vspace{-2em}

\thispagestyle{first}


\setstretch{1.08}

\section{Introduction}

High-$T_c$ superconductivity in cuprate materials has attracted immense research interest ever since its discovery. \cite{Bednorz&Mueller1986-SC-LaBaCuO,Keimer-highTc-Review,Tsuei&Kirtley2000,XGWen2006-highTc-Review,Proust&Taillefer2019}  The two-dimensional (2D) Hubbard model \cite{Hubbard1967,Arovas&Raghu2022} on a square lattice is widely believed to capture the essential physics of these systems, particularly the effects of strong electron correlations. In recent years, extensive researches have been carried out to explore the ground-state (zero-temperature) properties of the Hubbard model \cite{Giamarchi1991-Hubbard-VMC,Dagotto1994-Hubbard-Review,Halboth2000-Hubbard-DMRG,Huscroft&Tahvildarzadeh2001-QMC,White2003-Hubbard-DMRG,Hager2005-Hubbard-DMRG,Macridin&Dazevedo2005-QMC,Maier2005-Hubbard-MC,Capone2006-Hubbard-DMFT,Scalapino&Scalapino2006,Chang2010-Hubbard-AFQMC,Sordi2012-Hubbard-DMFT,Yokoyama2012-Hubbard-VMC,Gull2013-Hubbard-Summary,Otsuki2014-Hubbard-DMFT,2DHubbard-Benchmark,Deng2015-Hubbard-DiagMC,Corboz2016-extrap,Tocchio2016-Hubbard-GfMC,Zheng2016-DMET-Hubbard,NNHubbard-Conclusive,Zhao2017-Hubbard-VMC,Darmawan&Imada2018,Vanhala2018-Hubbard-DMFT,Corboz2019-Hubbard-PEPS,HCJiang2019-Hubbard-DMRG,Robinson&Tsvelik2019,Jiang&Jiang2020,Qin&Problem2020,White&Schollwoeck2020-DMRG-Hubbard-PlaquettePairing,Dong&Gull2022,Dong&Millis2022,Qin&Gull2022,Xu&Zhang2022,Jiang&White2023-3to1-Hubbard,Xiao&Zhang2023,Jiang&Devereaux2024-Hubbard-ehdoped,Schollwoeck&White2024-Hubbard-ehdoped,Sousa-junior&DosSantos2024,Zhang&VonDelft2025} or its simplified counterpart, $t$-$J$ model \cite{White&Scalapino1998,Hu&Sorella2012,Corboz2014-tJ-PEPS,Dodaro&Kivelson2017,Jiang&Kivelson2018,Grusdt&Cirac2020-iPEPS-Correlator,Jiang&Weng2020,Jiang&Kivelson2021,JWLi2021-tJ-PEPS,SSGong2021-tJ-DMRG,STJiang&Scalapino&White2021-t1t2J,STJiang&Scalapino&White2022-tttJ,Wietek&Wietek2022,Jiang&Lee2023,Lu&Gong2023,Lu&Gong2024-tJ-DMRG,Lu&Weng2024-tJ-sign,Lu&Gong2025}. In particular, long-range or quasi-long-range pairing orders, indicative of robust superconductivity, have been observed in the presence of a positive next-nearest-neighbor hopping amplitude $t'$ \cite{White&Schollwoeck2020-DMRG-Hubbard-PlaquettePairing,SSGong2021-tJ-DMRG,STJiang&Scalapino&White2021-t1t2J,STJiang&Scalapino&White2022-tttJ,Jiang&Devereaux2024-Hubbard-ehdoped,Lu&Gong2024-tJ-DMRG,Lu&Weng2024-tJ-sign,Lu&Gong2025,Zhang&VonDelft2025}. This naturally raises the question of how much such superconducting behavior extends to higher temperatures.

However, finite-temperature studies of the Hubbard model remain challenging due to the limitations of traditional numerical techniques. For example, the finite-temperature Lanczos method is constrained to small system sizes \cite{Jaklic&Prelovsek1994, Jaklic&Prelovsek1996, Jaklic&Prelovsek2000}, while Quantum Monte Carlo (QMC) is plagued by the notorious sign problem at finite doping \cite{Chen&Jarrell2013, He&Zhang2019-AFQMC-sign}. In this context, thermal tensor network algorithms have emerged as promising alternatives. \cite{White&White2009-METTS,Stoudenmire&White2010-METTS,Li&Su2011-LinearizedTRG,Czarnik2012-PEPS-finiteT-first,Czarnik2014-PEPS-finiteT-fermionic,Czarnik2015-PEPS-finiteT-variational,Czarnik&Dziarmaga2016,Li&Weichselbaum2018-XTRG,Chen&Li2019-XTRG,Czarnik&Corboz2019,Czarnik&Corboz2019a,Li&vonDelft2019-Heisenberg-XTRG,Chen&vonDelft2021-Hubbard-XTRG,Wietek&Georges2021-METTS,Wietek&Stoudenmire2021-METTS,Lin&Shi2022-XTRG,Sinha&Dziarmaga2022,Li2022-tanTRG,Qu&Su&Li2024-ttJ-tanTRG}. In particular, the eXponential Tensor Renormalization Group (XTRG) \cite{Li&Weichselbaum2018-XTRG} method excels in performing exponentially rapid cooling of the system, albeit at the cost of relatively high computational complexity.

In this work, we apply the Controlled Bond Expansion (CBE) technique \cite{Gleis&VonDelft2022,Gleis&VonDelft2023,Li&VonDelft2024-TDVP} originally designed for Matrix Product State (MPS) methods such as Density Matrix Renormalization Group (DMRG) \cite{Gleis&VonDelft2023} and Time-Dependent Variational Principle \cite{Li&VonDelft2024-TDVP} to products of Matrix Product Operators (MPOs), thereby introducing an enhanced 1s\textsuperscript{+} XTRG algorithm.  By exploring an enlarged variational space, our scheme attains 2-site update accuracy at a much lower computational complexity, delivering up to a 50\% acceleration for Hubbard-like systems.  This improvement enables cooling down to $T/t \!=\! 1/256$ (approximately $20\,\mathrm{K}$ for $t\!=\!0.3\!\sim\!0.5\,\mathrm{eV}$) and facilitates quantitative comparison with zero-temperature infinite Projected Entangled Pair State (iPEPS) results \cite{Zhang&VonDelft2025}. Pairing correlations are found to be enhanced at positive next-nearest neighbor (NNN) hopping ratio $t'/t$, and the pseudogap behavior, together with a possible Nagaoka polaron, is identified through the temperature dependence of the spin susceptibility. Finally, we generate and validate a comprehensive dataset of snapshots spanning the underdoped, intermediate-doped, and overdoped regimes across high, medium, and low temperatures. This dataset offers valuable resources for future Artificial Intelligence (AI)-driven analyses \cite{Bohrdt&Knap2019,Grusdt2019-Hubbard-StringPattern} of the Hubbard model and for comparison or calibration of cold-atom experiments \cite{Mazurenko&Greiner2017,Chiu&Greiner2018,Grusdt2019-Hubbard-StringPattern,Koepsell2019-Ultracold-FermiHubbard,Salomon&Gross2019,Koepsell2020-Ultracold-tech,Chen&vonDelft2021-Hubbard-XTRG,Koepsell2021-Ultracold-Polaron,Sompet&Bloch2022,Hirthe&Hilker2023,Xu&Greiner2023,Chalopin&Bloch2024,Pasqualetti&Folling2024,Bourgund&Hilker2025,Chalopin&Hilker2025,Xu&Greiner2025}.

\section{Model and Method}

In this paper, we consider the 2D $\tnn$-$\tnnn$ Hubbard model on an 8×8 square lattice with periodic boundary conditions (PBC) on the $y$ direction and open boundary conditions (OBC) on the $x$ direction unless stated otherwise. The Hamiltonian is defined as
\begin{equation}\label{Hamiltonian}
    \mathcal{H} = -\sum_{i,j,\sigma} t_{ij} \left[\, c^\dagger_{i\sigma} c_{j\sigma} + \text{h.c.} \,\right] + U\sum_i n_{i\uparrow} n_{i\downarrow}.
    \vspace{-4pt}
\end{equation}
\noindent Here, $t_{ij} = \tnn$ or $\tnnn$ for nearest neighbor (NN) or NNN hopping amplitude, respectively, and zero otherwise; $U$ measures the on-site Coulomb repulsion. Throughout this paper, we use $U/\tnn = 10$, as established to be realistic for cuprate materials \cite{Hirayama2018-Hubbard-parameter,Hirayama2018-Hubbard-electronic}, and set $\tnn=1$ for convenience.

A comparison of band structure between cuprate superconductors and the Hubbard model suggests a positive $\tnnn/\tnn$ ratio for electron-doped cuprates and a negative $\tnnn/\tnn$ for hole-doped cuprates \cite{Hirayama2018-Hubbard-parameter,Hirayama2018-Hubbard-electronic}. However, we note that this identification leads to inconsistencies between model predictions and experimental observations for crucial order parameters, as reported previously \cite{Li2022-tanTRG,Zhang&VonDelft2025,Wang&Devereaux2025-edoped-Hubbard} and further corroborated in the present work (see Section~\ref{sec:PairCorr} on pair-pair correlations).

The XTRG algorithm \cite{Li&Weichselbaum2018-XTRG} provides an efficient framework for constructing the thermal density matrix $\rho = \exp(-\beta\,\mathcal{H})$ of a quantum many-body system in the form of a MPO, where $\beta = 1/T$ is the inverse temperature. Specifically, an initial density matrix $\rho(\beta_0)$ is prepared at a high temperature (small $\beta_0\! = \mkern-2mu2^{-n_0}$) via the series expansion
\begin{equation}
    \rho(\beta_0) = \mathbbm{1} - \beta_0 \mathcal{H} + \frac{1}{2!} \beta_0^2 \mathcal{H}^2 + \cdots.
\end{equation}
The density matrix at half the temperature, or equivalently, at inverse temperature $2\beta$, is then obtained by squaring the density matrix at $\beta$:
\begin{equation}
    \rho(2\beta) = \exp\{ -2\beta \mathcal{H} \} = \rho(\beta)\cdot\rho(\beta).
\end{equation}
After $n$ iterations, one reaches an inverse temperature of $2^n\beta_0$, realizing an exponentially fast cooling of the system. The principal computational bottleneck in XTRG thus lies in the efficient MPO-MPO product.

Analogous to the DMRG, XTRG supports both 1-site and 2-site update schemes for MPO-MPO products. The 1-site update is computationally cheaper but limited by a restricted variational space. The 2-site update overcomes this limitation by exploring a larger variational manifold, at the cost of much higher computational complexity.

The Controlled Bond Expansion (CBE) technique \cite{Gleis&VonDelft2022,Gleis&VonDelft2023,Li&VonDelft2024-TDVP} offers a balanced compromise: by modestly enlarging the variational space, CBE achieves near 2-site update accuracy while maintaining a cost closer to that of the 1-site update. In the following section, we demonstrate how the CBE technique can be adapted to the MPO-MPO product, leading to a 1s\textsuperscript{+} scheme that significantly accelerates the XTRG algorithm.

Our simulations employ the state-of-the-art QSpace tensor library \cite{Weichselbaum2012-QSpace,Weichselbaum2012-QSpace-XSymbols,Weichselbaum&Weichselbaum2024,osqspacev4}, which preserves the $\mathrm{U}(1)_{\mathrm{charge}}\times\mathrm{SU}(2)_{\mathrm{spin}}$ symmetry in the MPO representation of the thermal density matrix. This allows us to retain up to $D^* = 1500$ multiplets (approximately $D\simeq 4000$ individual states), ensuring sufficient convergence of the XTRG algorithm.

\section{\texorpdfstring{1s\textsuperscript{+} MPO-MPO Product}{1s+ MPO-MPO Product}}

The MPO-MPO product constitutes a fundamental, and often the most computationally demanding, kernel in the XTRG algorithm. Both computing powers of $\mathcal{H}$ in the series expansion of $\rho(\beta_0)$ and performing the cooling step entail repeated MPO-MPO multiplications. A naive contraction of two MPOs with bond dimension $D$ yields an MPO of bond dimension $D^2$; compressing it back to $D$ via a straightforward singular value decomposition (SVD) incurs a cost of $\mathcal{O}(D^6)$. This overhead naturally motivates the development of variational compression schemes.

In the variational approach, one posits an ansatz for the target MPO and optimizes its proximity to the exact product. Concretely, given MPOs $A$ and $B$ with product $A \cdot B$, we seek an MPO $C$ that minimizes the squared Frobenius norm of the discrepancy
\vspace{-0.2em}
\begin{equation}
    \| C - A \cdot B \|_F^2 = C^\dagger C - C^\dagger (A \cdot B) + \text{h.c.} + \text{const.},
\end{equation}
where $\cdot$ denotes the MPO-MPO product. The optimization proceeds by sweeping along the chain and alternately updating one tensor at a time (1-site update, or 1s) or two adjacent tensors (2-site update, or 2s), while keeping all others fixed. The 1-site update requires solving the linear system
\vspace{-0.3em}
\begin{equation}
    \frac{\partial C^\dagger C}{\partial C_i^*} = \frac{\partial C^\dagger (A \cdot B)}{\partial C_i^*} 
    \vspace{-0.3em}
\end{equation}
iteratively for $i = 1, 2, \ldots, L$, where $L$ is the length of the MPO chain. For canonicalized MPOs, the left-hand side collapses to the tensor $C_i$ to be updated (denoted as 1s below). Diagrammatically, the 1-site update reads
\vspace{-0.7em}
\begin{equation}
    \vcenter{\hbox{\includegraphics[scale=0.5]{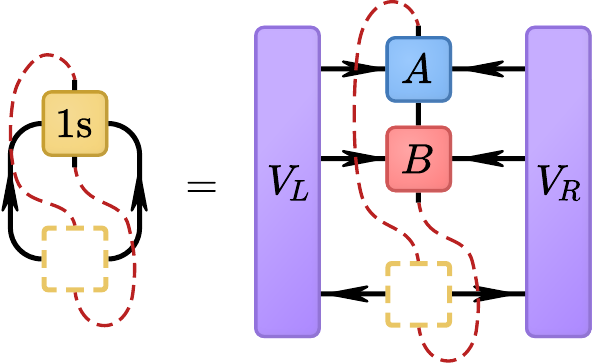}}}
\end{equation}
To circumvent the clutter introduced by twisty lines for physical indices, we streamline the notation by re-arranging the physical indices as follows
\vspace{-0.7em}
\begin{equation}
    \label{eq:OSUpdate}
    \vcenter{\hbox{\includegraphics[scale=0.5]{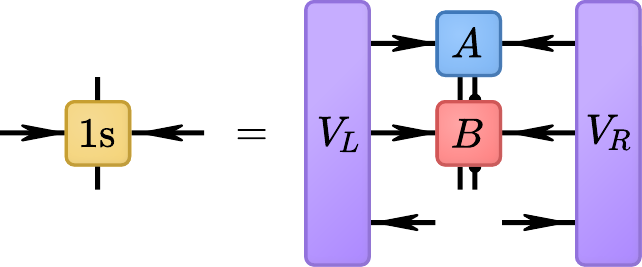}}}
\end{equation}
where the environment tensors $V_L$ and $V_R$ are pre-computed and held fixed during the 1-site update. The semicircles at the contact points between the physical index and tensor $B$ indicate that the physical index passes through without contraction.

A prevalent flaw of the 1-site update is its inability to enlarge the variational space in the presence of symmetry constraints. As implied by Eq.~\eqref{eq:OSUpdate}, the updated tensor remains confined to the symmetry sector of the original tensor. This constraint can be alleviated by the 2-site update, which simultaneously optimizes two neighboring tensors, i.e.
\begin{equation}
    \frac{\partial C^\dagger C}{\partial (C_i C_{i+1})^*} = \frac{\partial C^\dagger (A \cdot B)}{\partial (C_i C_{i+1})^*}\, .
\end{equation}
Diagrammatically, the 2-site update reads
\begin{equation}
    \vcenter{\hbox{\includegraphics[scale=0.5]{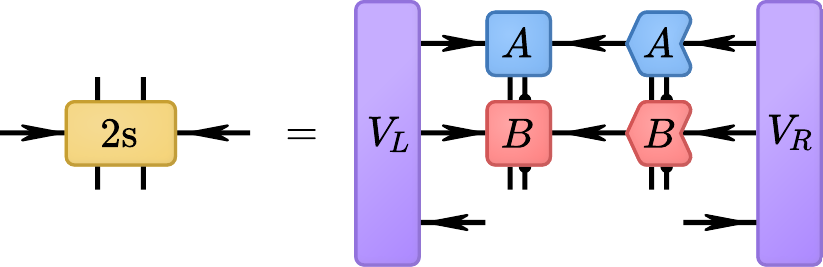}}}
\end{equation}
The meaning of the polygon-shaped tensor will be clarified later. The optimized 2-site tensor is then decomposed via an SVD to produce the updated MPO tensors, i.e.
\begin{equation}
    \label{eq:TSUpdate_SVD}
    \vcenter{\hbox{\includegraphics[scale=0.5]{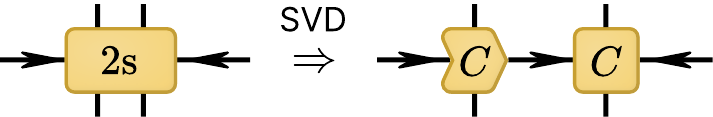}}}
\end{equation}
Note that a full SVD in Eq.~\eqref{eq:TSUpdate_SVD} generates in total $Dd^2$ components, where $d$ is the physical Hilbert-space dimension, whereas only the leading $D$ components are ultimately retained. Consequently, most components are computed only to be discarded, squandering computational resources. Our 1s\textsuperscript{+} scheme mitigates this inefficiency by injecting only a small fraction of \emph{additional components}, thereby enlarging the accessible variational space without incurring the full cost of a 2-site update. 

In the spirit of \cite{Gleis&VonDelft2022}, a full MPO tensor admits a decomposition into a discarded-space component and a kept-space component
\begin{equation}
    \vcenter{\hbox{\includegraphics[scale=0.5]{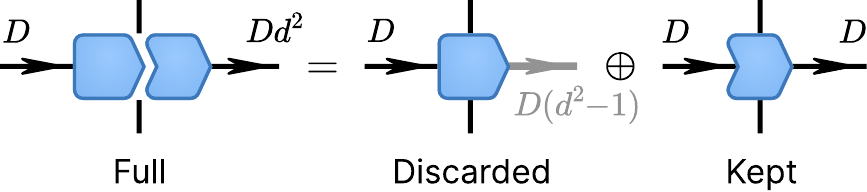}}}
\end{equation}
\indent In our notation, tensors are depicted as (composite) polygons. The discarded component is denoted by a pentagon with a wedge, and the kept component by a polygon with both a wedge and a notch. Gray lines indicate components from the discarded space. The protruding face indicates the direction toward the orthogonality center of the MPO, or equivalently, the opposite direction to the normalization. In the example above, the MPO tensors are left-normalized, placing the orthogonality center to the right of the tensor. Directions of the arrows encode the flow of quantum numbers, following the convention of \cite{Weichselbaum&Weichselbaum2024}.

Similar to the CBE technique \cite{Gleis&VonDelft2023}, the discrepancy between the variational spaces accessible to the 1-site and 2-site updates can be identified by the following construction
\begin{equation}
    \vcenter{\hbox{\includegraphics[scale=0.5]{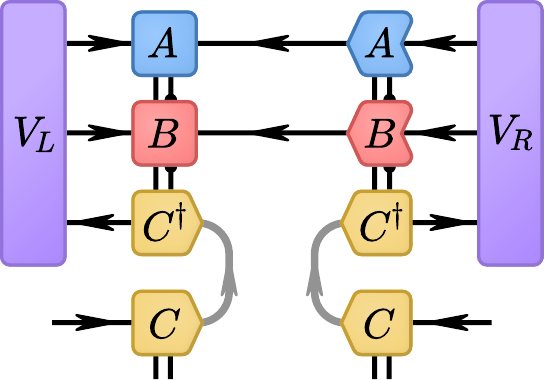}}}
\end{equation}
\indent The 2-site discarded-space tensor spans the variational sector accessible to the 2-site update but excluded from the 1-site manifold. An exact construction of this tensor would incur a computational cost $\mathcal{O}(D^4d^4)$ comparable to a full 2-site update. The dominant cost resides in the contraction of the left and right part of the above 2-site tensor. Therefore, for practical purposes, we instead distill its principal components by performing the following SVD
\begin{equation}
    \vcenter{\hbox{\includegraphics[scale=0.5]{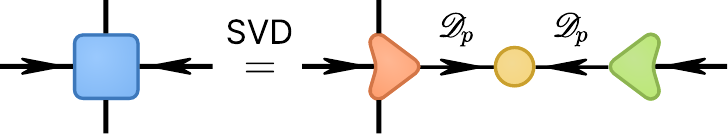}}}
\end{equation}
and retain only the leading $\mathscr{D}_{\mkern-2mu p}$ components. Here, $\mathscr{D}_{\mkern-2mu p}$ serves as a hyperparameter that controls the fidelity of the 2-site discarded-space approximation. In our simulations, we choose $\mathscr{D}_{\mkern-2mu p} = \text{round}(\sqrt{D})$ to balance accuracy against computational overhead.
With this construction, the green tensor acts as a projector that compresses the bond dimension further down to $\mathscr{D}_{\mkern-2mu p}$. The principal components of the 2-site discarded-space tensor are then extracted by inserting these projectors as
\begin{equation}
    \label{eq:Complement}
    \vcenter{\hbox{\includegraphics[scale=0.5]{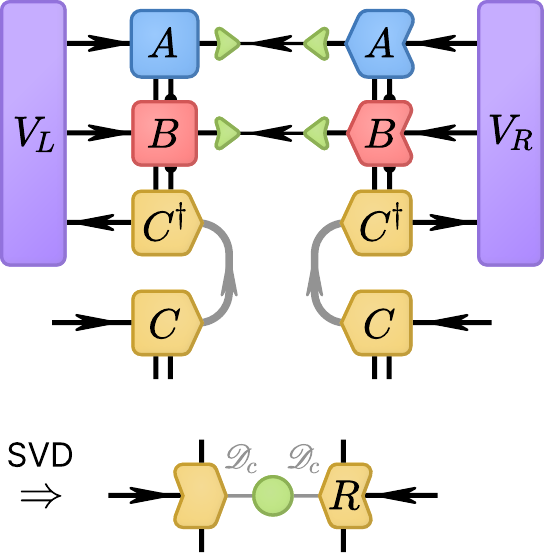}}}
\end{equation}
\indent We emphasize that this procedure yields only an \emph{approximation} to the principal components. In practice, exact identification of the principal components are unnecessary: the goal is to efficiently isolate the principal subspace that complements the 1-site update, after which subsequent optimization sweeps would converge to the desired solution.

We now outline the ensuing steps for a left-to-right optimization sweep; the right-to-left sweep proceeds analogously. The 2-site discarded-space tensor is decomposed and truncated to $\mathscr{D}_{\mkern-2mu c}$ via an SVD (see Eq.~\eqref{eq:Complement}), yielding the isometry $R$ tailored for the left-to-right sweep. The tensor $R$ thus serves as the supplement to the MPO tensor on the right. In our computations, we set $\mathscr{D}_{\mkern-2mu c} = \text{round}(\sqrt{D})$. Next, the left supplement tensor, $L$, is obtained by
\begin{equation}
    \vcenter{\hbox{\includegraphics[scale=0.5]{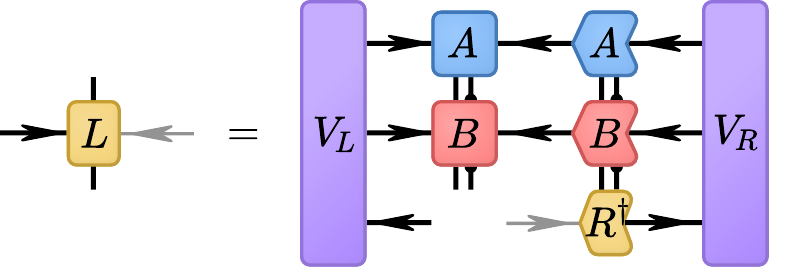}}}
\end{equation}
\indent Tensors $L$ and $R$ thus offers the \emph{additional components} that augment the variational space. Equipped with these ingredients, we assemble the expanded MPO tensors $L^+$ and $R^+$ as follows
\begin{center}
\vspace*{-3em}
\begin{equation}
    \vcenter{\hbox{\includegraphics[scale=0.5]{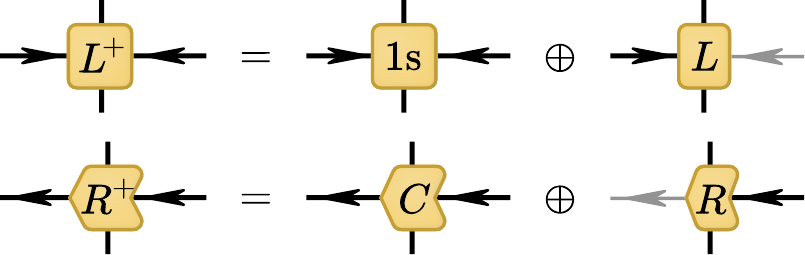}}}
\end{equation}
\end{center}
\indent Contracting $L^+$ and $R^+$ produces a tensor analogous to the 2-site object in Eq.~\eqref{eq:TSUpdate_SVD}, albeit with substantially fewer auxiliary components. The updated MPO tensors are then obtained by applying an SVD and truncating to the leading $D$ components
\begin{equation}
    \vcenter{\hbox{\includegraphics[scale=0.5]{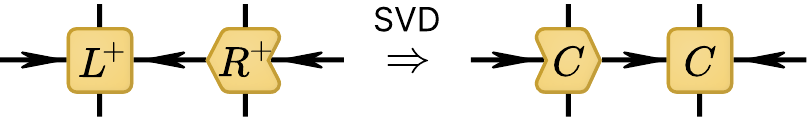}}}
\end{equation}
\indent The canonical 2-site update exhibits a computational complexity $\mathcal{O}(D^4d^4)$, where $d$ denotes the physical dimension of the MPO tensors; this scaling is dominated by constructing the updated 2-site tensors. By contrast, the 1s\textsuperscript{+} scheme reduces the complexity to the 1-site level, i.e., $\mathcal{O}(D^4d^2)$, thereby shifting the computational bottleneck to constructing the environmental tensors $V_L$ and $V_R$. Consequently, the overall complexity of the 1s\textsuperscript{+} scheme is $\mathcal{O}(D^4d^2 + D^4d^3)$. For Hubbard-like systems, the physical dimension $d=4$, rendering a maximum $4\times$ speedup relative to the 2-site update. In practice, we observe speedups of up to $50\%$ for the Hubbard model, owing to the overheads from the additional operations. The realized acceleration also depends on the specific choice of $\mathscr{D}_{\mkern-2mu p}$ and $\mathscr{D}_{\mkern-2mu c}$.

\section{Benchmarks}

\begin{figure}[htp!]
    \vspace{0.1em}
    \centering
    \includegraphics[scale=0.55]{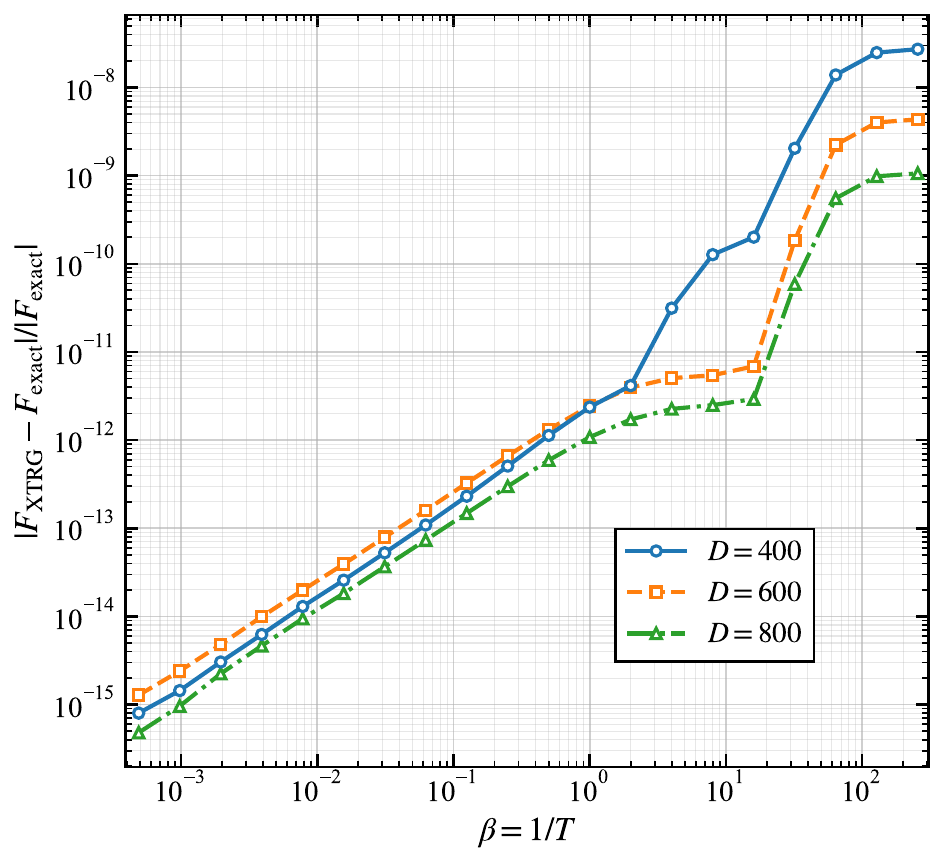}
    \begin{minipage}{0.45\textwidth}
    \caption{The error of the free energy $F$ of the free-fermion model relative to the exact value as a function of the inverse temperature $\beta$, obtained via the 1s\textsuperscript{+} XTRG algorithm with bond dimension $D = 400, 600, 800$, respectively.}
    \label{fig:FreeFerm}
    \end{minipage}
\end{figure}

In this section, we present benchmarks of the 1s\textsuperscript{+} XTRG algorithm on an analytically solvable free-fermion model. We adopt the following Hamiltonian defined on a one-dimensional open-boundary chain of length $L$:
\begin{equation}
    \mathcal{H} = -t\sum_{i=1}^{L-1} \left[\, c^\dagger_i c_{i+1} + c^\dagger_{i+1} c_i \,\right].
\end{equation}
This Hamiltonian is exactly diagonalizable, yielding eigenvalues $\epsilon_k = -2t\cos(k\pi / (L+1))$ for $k = 1, \ldots, L$. Accordingly, the free energy can be computed as
\begin{equation}
    F = -T \, \sum_{k=1}^{L} \ln (1 + e^{-\beta\epsilon_k})\, .
\end{equation}
\indent In our benchmarks, we initialize the density matrix at a high temperature of $\beta = 2^{-12}$ and perform $20$ 1s\textsuperscript{+} XTRG iterations to cool the system down to $\beta = 2^8$. Fig.~\ref{fig:FreeFerm} displays the relative error of the free energy $F$ as a function of the inverse temperature $\beta$, computed via the 1s\textsuperscript{+} XTRG algorithm with bond dimensions $D = 400, 600, 800$, respectively. The numerics indicate that the 1s\textsuperscript{+} XTRG algorithm achieves accuracy comparable to the 2-site update scheme \cite{Li&Weichselbaum2018-XTRG}, while delivering a substantial speedup.

\begin{figure*}[htp!]
    \centering
    \includegraphics[scale=0.54]{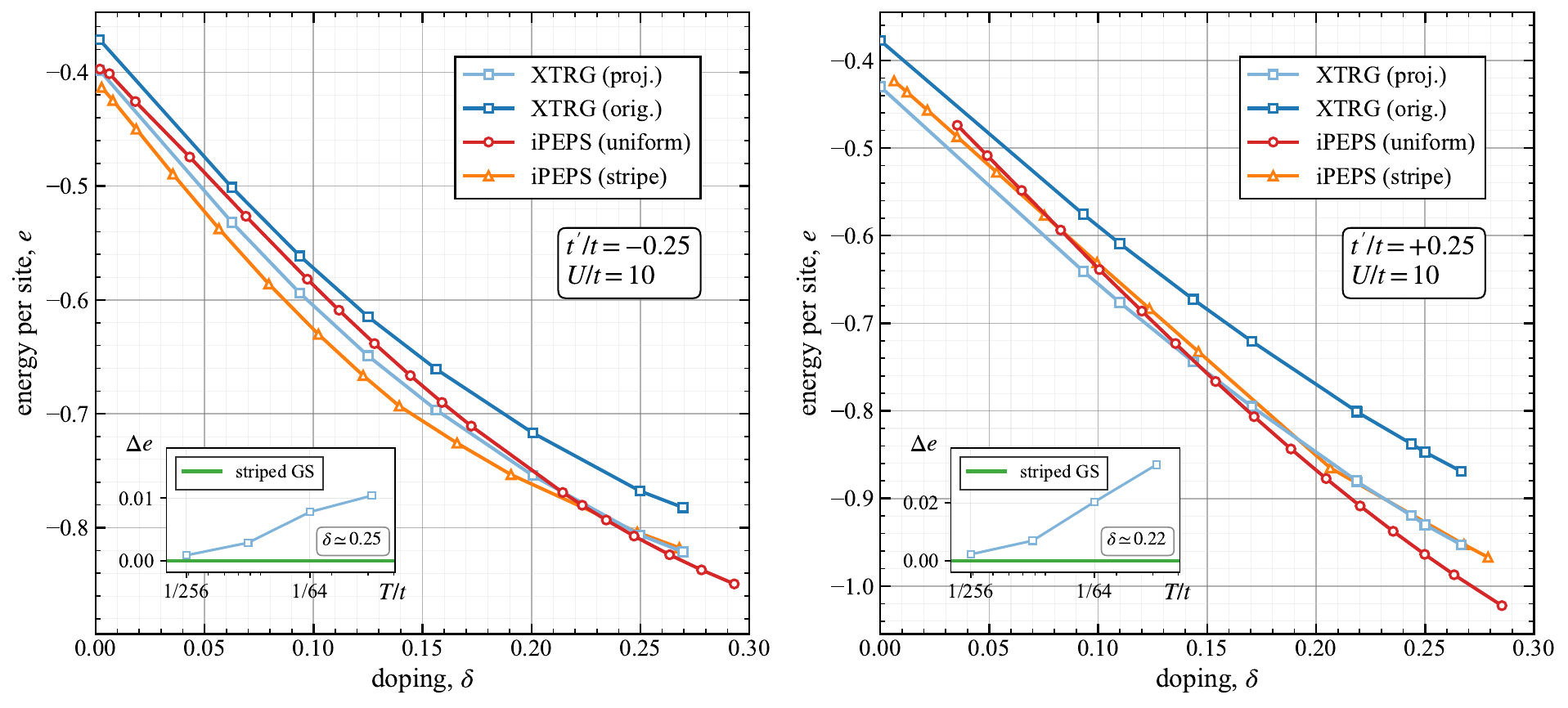}
    \begin{minipage}{0.9\textwidth}
    \caption{The ground state energy per site (red for uniform and orange for striped states) obtained from iPEPS simulations, and the energy at $T/t = 1/256$ (blue) obtained from the 1s\textsuperscript{+} XTRG algorithm. The iPEPS ground states are acquired on an infinite lattice with 4×2 (uniform) or 8×2 (striped) supercell and the XTRG density matrices are generated on an 8×8 lattice with PBC on the $y$ direction. The light blue curves mark the projected energies for the XTRG density matrices with PBC on both directions. The insets show the difference $\Delta e$ between the projected XTRG energies and the striped iPEPS ground state (GS) energies with respect to temperature for two representative doping levels.}
    \label{fig:XTRG_iPEPS}
    \end{minipage}
\end{figure*}

\section{Energetics of the Hubbard Model}

In this section, we juxtapose the energetics of the $t$-$t'$ Hubbard model obtained from an iPEPS ground-state search with those from XTRG cooling. Our enhanced 1s\textsuperscript{+} XTRG algorithm enables a cooling of the system down to a temperature of $T/t \!=\! 1/256$, sufficiently low to warrant a comparison with zero-temperature ground state studies.

The iPEPS ground state search is performed on an infinite lattice, achieved by PBC on both $x$ and $y$ directions, with a 4×2 (uniform) or 8×2 (striped) supercell, following \cite{Zhang&VonDelft2025}. Ground states of distinct characteristics are targeted by constraining the spin symmetry of the tensor network: imposing $\mathrm{SU}(2)_{\text{spin}}$ yields a uniform state, whereas enforcing $\mathrm{U}(1)_{\text{spin}}$ produces a striped state. The uniform states retain $D^*=7$ multiplets, equivalent to $D=12$ individual states; for consistency, the striped states are fixed to a bond dimension of $D=12$.

The XTRG runs start from a high temperature of $\beta = 2^{-12}$ on an 8×8 lattice with PBC along the $y$ direction, and the system is cooled down to $\beta = 2^8$ via $20$ 1s\textsuperscript{+} XTRG iterations. Throughout, we preserve $\mathrm{U}(1)_\text{charge}\times\mathrm{SU}(2)_{\text{spin}}$ symmetry in the MPO representation of the thermal density matrix, allowing us to retain up to $D^* = 1500$ multiplets (approximately $D\simeq 4000$ individual states). We consider only the $\mathrm{SU}(2)_{\text{spin}}$ symmetry here since continous symmetry breaking is precluded at finite temperature by the Mermin-Wagner theorem \cite{Mermin&Wagner1966,Hohenberg&Hohenberg1967}.

The iPEPS ground states and XTRG density matrices are defined on lattices with differing boundary conditions, which in turn modifies the count of kinetic terms in the Hamiltonian. Specifically, an 8×8 lattice with PBC along both $x$ and $y$ directions contains $128$ nearest neighbor (NN) hopping terms and $128$ next-nearest neighbor (NNN) hopping terms, whereas the same lattice with PBC only along the $y$ direction comprises $120$ NN terms and $112$ NNN terms. To reconcile these discrepancies, we posit bulk-averaged energies for the missing boundary contributions. This leads to a projected (kinetic) energy with
\begin{equation}
    \frac{e^\text{kin}_\text{proj}}{e^\text{kin}_\text{orig}} = \frac{128\cdot e_\text{NN} + 128\cdot e_\text{NNN}}{120\cdot e_\text{NN} + 112\cdot e_\text{NNN}},
\end{equation}
where $e_\text{NN}$ and $e_\text{NNN}$ denote the mean energies of the NN and NNN hopping terms, respectively. This projected energy furnishes a more commensurate reference for comparison with the iPEPS results.

Fig.~\ref{fig:XTRG_iPEPS} displays the ground state energy (red for uniform and orange for striped states) obtained from iPEPS simulations, alongside the energy at $T/t = 1/256$ (blue) derived from the 1s\textsuperscript{+} XTRG algorithm. The light blue curves indicate the projected energies for the XTRG density matrices with PBC along both directions.

For both $\tnnn/\tnn = -0.25$ and $\tnnn/\tnn = +0.25$, the projected XTRG energies align well with the iPEPS ground state energies, showing a similar overall doping dependence. In particular, they lie quite close to the energies of the striped states, which may be attributed to stripe formation induced by the OBC along the $x$ direction in the XTRG simulations. This agreement indicates that our XTRG algorithm attains sufficient purity upon approaching the ultra-cold regime, i.e. the density matrix $\rho\simeq|\text{GS}\rangle\langle\text{GS}|$ at lowest temperature, where $|\text{GS}\rangle$ is the ground state.

\begin{figure*}
    \centering
    \includegraphics[scale=0.58]{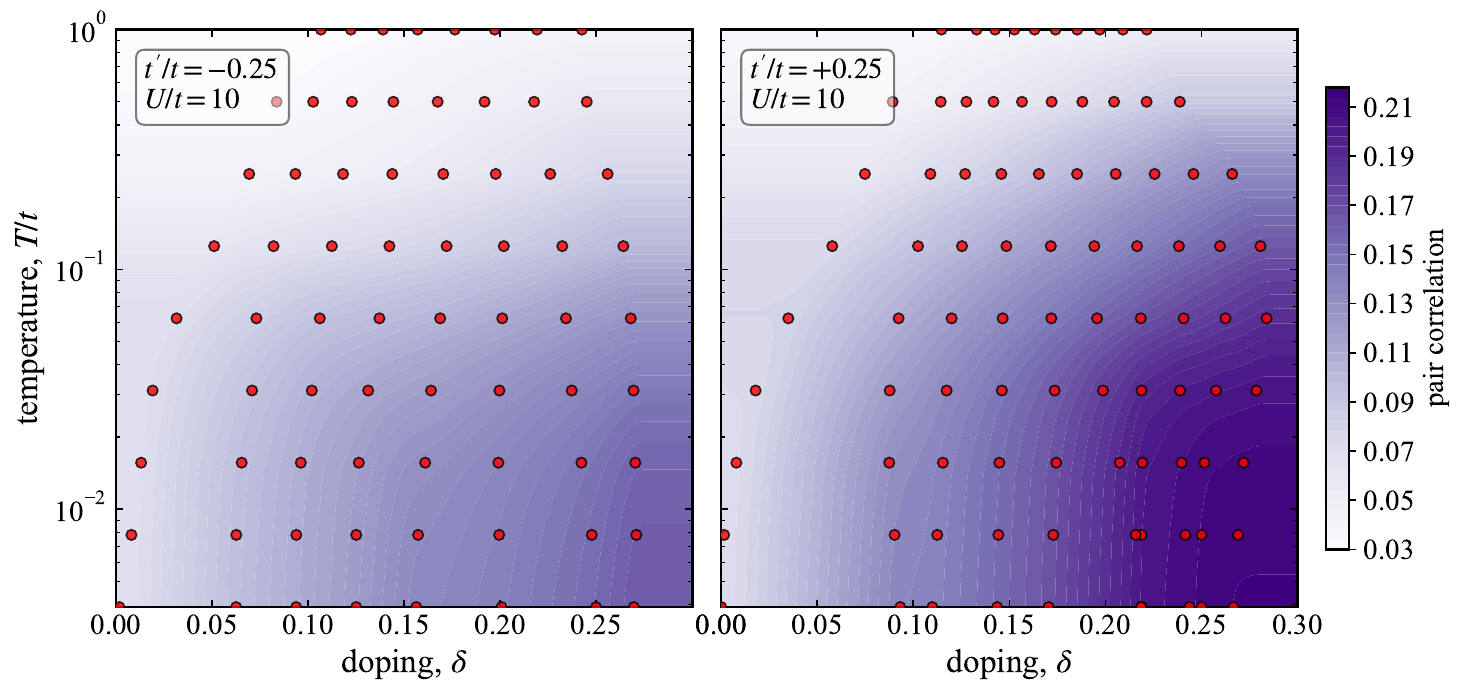}
    \begin{minipage}{0.9\textwidth}
    \caption{The pair correlation indicator as a function of doping $\delta$ and temperature $T/t$, obtained from the 1s\textsuperscript{+} XTRG for $\tnnn/\tnn = -0.25$ (left) and $\tnnn/\tnn = +0.25$ (right). Red circles mark the positions with exact data, and the colored contour map displays the interpolation between the data points. The pairing correlation is found strengthened at large doping, low temperature and a positive $\tnnn/\tnn$ ratio.}
    \label{fig:PairCorr}
    \end{minipage}
\end{figure*}

\section{Pair-Pair Correlations}
\label{sec:PairCorr}

In this section, we investigate the dependence of pairing correlation, which signifies the strength of superconductivity, on doping, temperature and the sign of $\tnnn/\tnn$. Rigorously speaking, anomalous local pairing amplitudes are forbidden by the one-dimensionality of the tensor network together with the conserved $\mathrm{U}(1)_\text{charge}$ symmetry. We therefore focus on the pair-pair correlations, which quantify the propensity for pair propagation.

Here, we primarily examine the nearest neighbor singlet pairing correlation
\begin{equation}
    \rho_S(\mathbf{r},\bm{\alpha}|\mathbf{s},\bm{\beta}) = \langle \Delta^\dagger_{\mathbf{r},\mathbf{r} + \bm{\alpha}} \Delta_{\mathbf{s},\mathbf{s} + \bm{\beta}} \rangle,
\end{equation}
where the singlet pairing operator $\Delta_{\mathbf{r},\mathbf{r} + \bm{\alpha}}$ is defined as
\begin{equation}
    \Delta_{\mathbf{r},\mathbf{r} + \bm{\alpha}} = c_{\mathbf{r},\uparrow} c_{\mathbf{r} + \bm{\alpha},\downarrow} - c_{\mathbf{r},\downarrow} c_{\mathbf{r} + \bm{\alpha},\uparrow},
\end{equation}
and $\bm{\alpha} = \mathbf{x},\mathbf{y}$ denotes the horizontal or vertical unit vector. Following \cite{Wietek&Wietek2022,Baldelli&Wietek2024}, we zero-out the contributions whenever $\mathbf{r}$, $\mathrm{r} + \bm{\alpha}$, $\mathbf{s}$, or $\mathbf{s} + \bm{\beta}$ are not all distinct. 

The correlator $\rho_S(\mathbf{r},\bm{\alpha}|\mathbf{s},\bm{\beta})$, viewed as a matrix with composite indices $(\mathbf{r},\bm{\alpha})$ and $(\mathbf{s},\bm{\beta})$, can be diagonalized, and one can take the dominant eigenvalue as a scalar indicator of the pairing strength \cite{Penrose&Onsager1956,Wietek&Wietek2022,Baldelli&Wietek2024}.

Figure~\ref{fig:PairCorr} presents the dominant eigenvalue of $\rho_S$ as a function of doping $\delta$ and temperature $T/t$, obtained via the 1s\textsuperscript{+} XTRG for $\tnnn/\tnn = -0.25$ (left) and $\tnnn/\tnn = +0.25$ (right). The colored contour map shows the interpolation between discrete data points marked by the red circles.

We observe from Fig.~\ref{fig:PairCorr} that pairing correlations strengthen with increasing doping $\delta$ and decreasing temperature $T/t$, consistent with the general trends seen in iPEPS ground state simulations \cite{Zhang&VonDelft2025} and in cuprate phenomenology. Moreover, the correlations are systematically larger for positive $\tnnn/\tnn$, in line with various ground-state or finite-temperature studies of the $t$–$J$ \cite{SSGong2021-tJ-DMRG,STJiang&Scalapino&White2021-t1t2J,STJiang&Scalapino&White2022-tttJ,Lu&Gong2024-tJ-DMRG,Lu&Weng2024-tJ-sign,Lu&Gong2025,Qu&Su&Li2024-ttJ-tanTRG} and Hubbard models \cite{White&Schollwoeck2020-DMRG-Hubbard-PlaquettePairing,Jiang&Devereaux2024-Hubbard-ehdoped,Zhang&VonDelft2025,Li2022-tanTRG,Wang&Devereaux2025-edoped-Hubbard}. However, if one adopts the band-structure-based identification that for cuprate superconductors $\tnnn/\tnn\!>\!0$ or $<\!0$ corresponds to electron or hole doping, respectively, our finding that pairing is stronger for $\tnnn/\tnn\!>\!0$ than for $\tnnn/\tnn\!<\!0$ would then be at odds with the experimental observations that pairing is weaker for electron- than hole-doped systems. This emphasizes the necessity for further investigations regarding the appropriate parameter settings or additional terms in the effective models \cite{STJiang&Scalapino&White2021-t1t2J,Xiang2009-ElectronDopedCuprates,Jiang&White2023-3to1-Hubbard,Wang&Devereaux2025-edoped-Hubbard}.

\begin{figure*}
    \centering
    \includegraphics[scale=0.58]{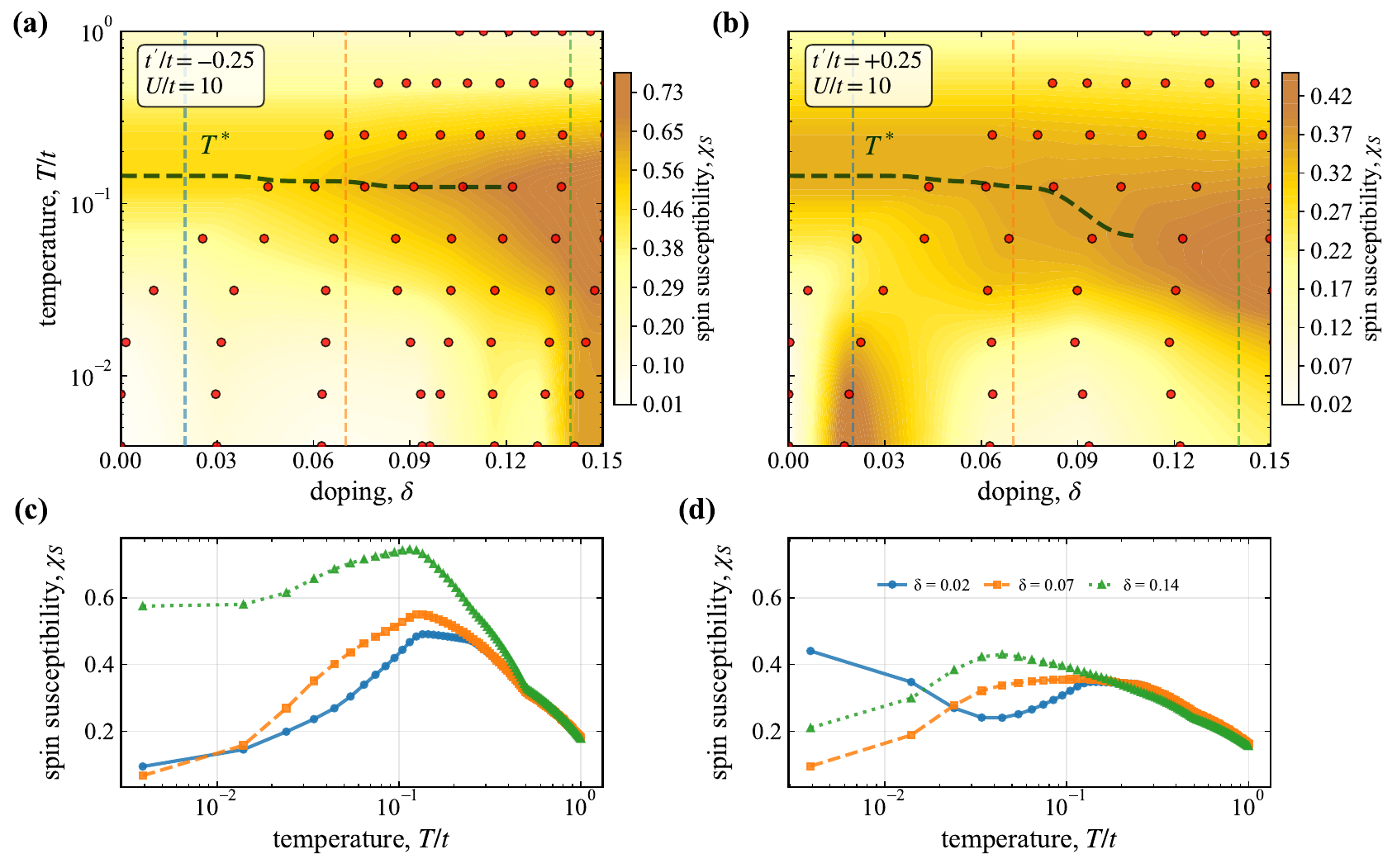}
    \begin{minipage}{0.9\textwidth}
    \caption{The spin susceptibility (a,b) as a function of doping $\delta$ and temperature $T/t$, and (c,d) as a function of temperature $T/t$ for three representative doping levels (via interpolation), obtained from the 1s\textsuperscript{+} XTRG for (a,c) $\tnnn/\tnn = -0.25$ and (b,d) $\tnnn/\tnn = +0.25$. Red circles mark the positions with exact data, and the colored contour map displays the interpolation between the data points. The thick dashed line indicates the locus of the susceptibility peak, signaling the onset temperature $T^*$ where pseudogap starts to develop.}
    \label{fig:Pseudogap}
    \end{minipage}
\end{figure*}

\vspace{-0.5em}
\section{Spin susceptibility and Pseudogap}
\vspace{-0.2em}

The pseudogap --- manifest as a partial depletion in the electronic density of states --- is a hallmark of underdoped cuprate superconductors. Its onset suppresses the low-energy spectral weight and, concomitantly, reduces the uniform spin susceptibility $\chi_S$. In practice, as temperature decreases, $\chi_S$ initially follows a Curie-Weiss-like increase, reaches a maximum at a characteristic scale $T^*$, and subsequently decreases as the pseudogap develops. The locus of this maximum thus provides a practical proxy for the pseudogap onset temperature $T^*$.

We compute the spin susceptibility by augmenting the Hamiltonian with a Zeeman term, $H_Z = -h \sum_i S^z_i$. The uniform susceptibility is then approximated as
\vspace{-0.3em}
\begin{equation}
    \chi_S = -\langle S^z \rangle / h
\end{equation}
for a sufficiently small external field $h\!=\!0.01$, where $S^z = \sum_i S^z_i / L$. To probe the response to a magnetic field, we relax the $\mathrm{SU}(2)_{\text{spin}}$ symmetry to $\mathrm{U}(1)_{\text{spin}}$ in these simulations, and retain a maximum bond dimension of $D=1500$.

Figure~\ref{fig:Pseudogap} compiles $\chi_S$ across doping $\delta$ and temperature $T/t$, obtained via the 1s\textsuperscript{+} XTRG for $\tnnn/\tnn = -0.25$ (left) and $\tnnn/\tnn = +0.25$ (right). The colored contour shading interpolates between discrete data points as indicated by red markers. The dashed curve tracks the ridge line of the susceptibility, identifying the onset scale $T^*$ at which the pseudogap begins to emerge.

From the figure, $T^*$ decreases monotonically with increasing $\delta$, with a more pronounced reduction for positive $\tnnn/\tnn$. These trends are broadly consistent with results for the $t$-$J$ model \cite{Qu&Su&Li2024-ttJ-tanTRG} and with observations in various cuprate compounds \cite{Loret&Sacuto2017,Boschini&Damascelli2020,Grissonnanche&Taillefer2023}. We note that the temperature axis is logarithmic, which visually attenuates the apparent variation of $T^*$ with doping. Besides, the overall magnitude of $\chi_S$ is smaller for positive $\tnnn/\tnn$, indicating a more enhanced pseudogap effect in this regime.

Moreover, for positive $\tnnn/\tnn$ we observe an anomalous low-temperature enhancement of $\chi_S$ near $\delta \simeq 0.02$. On an 8×8 lattice, this doping approximately corresponds to a single hole introduced into the half-filled system. The upturn can therefore plausibly be attributed to the formation of a Nagaoka polaron \cite{Nagaoka&Nagaoka1966,Tasaki&Tasaki1998,Lebrat&Greiner2024,Samajdar&Bhatt2024}, which engenders an underlying ferromagnetic tendency and exhibits a strong response to the external magnetic field.

\section{Sampling of the Density Matrix}

In this section, we elaborate a straight-forward sampling \cite{qu2024phasediagramdwavesuperconductivity} scheme for the thermal density matrix $\rho$ in the format of an MPO. This procedure can be used to generate an ensemble of snapshots (projections onto basis states of the Fock space) for each density matrix at any specified location in phase space. These ensembles can then be assembled into datasets for subsequent Artificial Intelligence (AI) analyses aimed at uncovering novel, non-trivial features of the Hubbard model.

We employ sequential sampling. Specifically, for site $i$, assuming all sites $\iota < i$ have already been sampled, the single-site reduced density matrix $\rho_i$ can be constructed via the partial trace:
\vspace{-0.2em}
\begin{equation}
    \vcenter{\hbox{\includegraphics[scale=0.5]{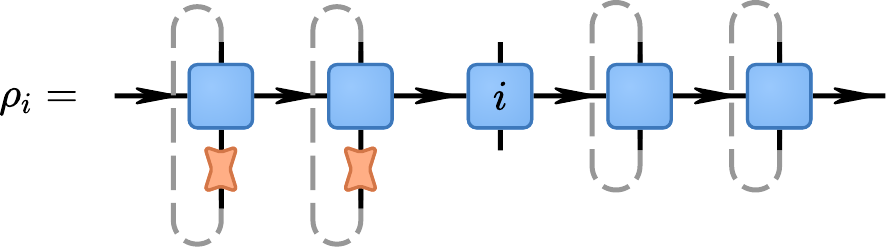}}}
    \vspace{-0.2em}
\end{equation}
where the local state projector is defined as
\vspace{-0.2em}
\begin{equation}
    \vcenter{\hbox{\includegraphics[scale=0.5]{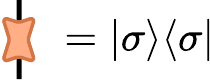}}}
    \vspace{-0.2em}
\end{equation}
with $\sigma$ denoting the sampled local state of the corresponding site. The diagonal entries of $\rho_i$ thus furnish the probability distribution for the local state $\sigma_i$ at site $i$. Sampling proceeds by drawing $\sigma_i$ according to this distribution and then repeating the procedure for the next site $i+1$.

In this work, we generate snapshot ensembles for the minimal Hubbard model ($t'=0$) on an 8×8 open-boundary lattice intended for future comparisons with modern ultra-cold atom experiments \cite{Koepsell2019-Ultracold-FermiHubbard,Koepsell2020-Ultracold-tech,Chen&vonDelft2021-Hubbard-XTRG,Koepsell2021-Ultracold-Polaron,Xu&Greiner2023,Pasqualetti&Folling2024,Xu&Greiner2025}. The thermal density matrix MPOs are produced using the 1s\textsuperscript{+} XTRG method across the doping range of $0\!<\!\delta\!<\!0.25$ and down to a lowest temperature of $T/t = 1/256$. To resolve the spin orientation, we relax the symmetry here from $\mathrm{SU(2)}_\text{spin}$ down to $\mathrm{U(1)}_\text{spin}$. Applying the sampling protocol described above, we draw $1000$ snapshots for each combination of doping and temperature.

\begin{figure}[htp!]
    \centering
    \includegraphics[scale=0.55]{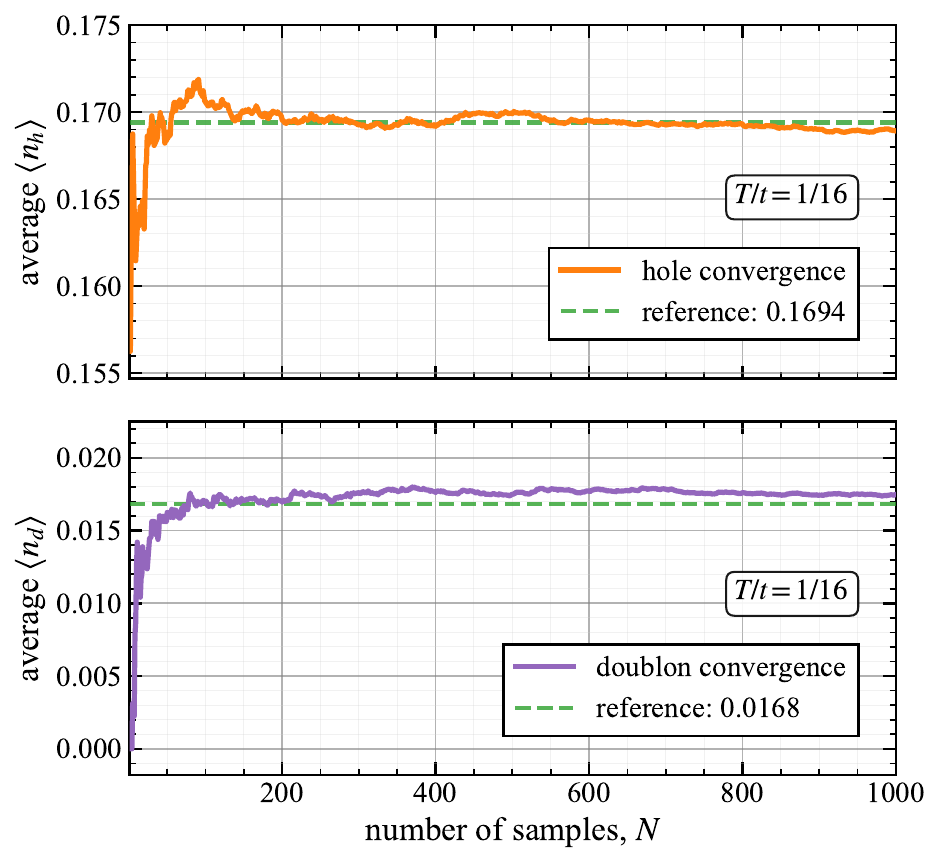}
    \begin{minipage}{0.45\textwidth}
        \caption{Convergence behavior of hole density $n_h$ and double occupancy $n_d$ as a function of sample size at $\delta\simeq 0.1694$ and $T/t = 1/16$. Green dashed lines indicate reference values extracted directly from the density matrix.}
        \label{fig:Conv}
    \end{minipage}
\end{figure}

Figure~\ref{fig:Conv} documents the convergence of the hole density $n_h$ and double occupancy $n_d$ as a function of sample size at $\delta\simeq 0.1694$ and $T/t = 1/16$. Green dashed lines indicate reference values extracted directly from the density matrix. We find that satisfactory convergence is achieved for $\gtrsim 200$ samples; accordingly, a sample size of $1000$ is sufficient to furnish a representative ensemble for the corresponding point in the phase space.

\section{Summary and Outlook}

In this work, we introduced an enhanced 1s\textsuperscript{+} XTRG algorithm that accelerates the cooling of the two-dimensional Hubbard model at finite temperature. By enlarging the accessible variational manifold, the method attains accuracy comparable to a two-site update while incurring computational cost near that of a one-site update, delivering speedups of up to $50\%$ for Hubbard-like systems. This improvement enables cooling down to $T/t \approx 0.004$ and supports direct comparisons with zero-temperature iPEPS simulations. We demonstrate that the projected XTRG energies and pairing-correlation characteristics are in close accord with iPEPS ground-state results, indicating that our XTRG approach achieves sufficient purity upon entering the ultra-cold regime.

Our finite-temperature simulations provide a systematic characterization of singlet pairing correlations and spin susceptibilities across a broad range of dopings and temperatures. We find that pairing correlations are enhanced at large doping, low temperature, and for positive $\tnnn/\tnn$; the latter trend accords with numerous prior numerical studies yet contrasts with behaviors observed in cuprate materials. The pseudogap onset temperature $T^*$ decreases with increasing doping for both signs of $\tnnn/\tnn$, in line with experimental observations. Finally, we identify an anomalous low-temperature enhancement of the spin susceptibility within a narrow underdoped regime, which may be attributable to the formation of a Nagaoka polaron.

In addition, we leveraged a sequential sampling scheme that generates snapshot ensembles for the minimal Hubbard model, each comprising $1000$ samples and shown to be statistically representative of the corresponding point in phase space. These snapshots furnish a data resource for analyzing thermal properties and correlations, opening the door to future AI-driven investigations of the Hubbard model. A comprehensive analysis of this dataset will be presented in a forthcoming publication.

\vspace{2em}
\section*{Acknowledgements}

We acknowledge fruitful discussions with Annabelle Bohrdt, Yuan Gao, Ming Huang, Qiaoyi Li, Wei Li, and Dai-Wei Qu.
This research was funded in part by the Deutsche Forschungsgemeinschaft under Germany's Excellence Strategy EXC-2111 (Project No.~390814868), and is part of the Munich Quantum Valley, supported by the Bavarian state government through the Hightech Agenda Bayern Plus.

\bibliography{library}

\end{document}